\documentclass[prl,twocolumn,showpacs,preprintnumbers,amsmath,amssymb,superscriptaddress]{revtex4}

\usepackage{graphicx}% Include figure files
\usepackage{dcolumn}% Align table columns on decimal point
\usepackage{bm}% bold math

\begin{document}

%\preprint{APS/123-QED}

\title{Quantum simulator for the Hubbard model with long-range Coulomb interactions using surface acoustic waves}
\author{Tim Byrnes}
\affiliation{National Institute of Informatics, 2-1-2
Hitotsubashi, Chiyoda-ku, Tokyo 101-8430, Japan}

\author{Patrik Recher}
\altaffiliation{Also at Institute of Industrial Science,
University of Tokyo, 4-6-1 Komaba, Meguro-ku, Tokyo 153-8505,
Japan} \affiliation{E. L. Ginzton Laboratory, Stanford University,
Stanford, CA 94305}
\author{Na Young Kim}
\affiliation{E. L. Ginzton Laboratory, Stanford University,
Stanford, CA 94305}
\author{Shoko Utsunomiya}
\affiliation{National Institute of Informatics, 2-1-2
Hitotsubashi, Chiyoda-ku, Tokyo 101-8430, Japan}
\author{Yoshihisa Yamamoto}
\affiliation{National Institute of Informatics, 2-1-2
Hitotsubashi, Chiyoda-ku, Tokyo 101-8430, Japan}
 \affiliation{E.
L. Ginzton Laboratory, Stanford University, Stanford, CA 94305}
\date{\today}% It is always \today, today,

\begin{abstract}
A practical experimental scheme for a quantum simulator of
strongly correlated electrons is proposed. Our scheme employs
electrons confined in a two dimensional electron gas in a
GaAs/AlGaAs heterojunction. Two surface acoustic waves are then
induced in the GaAs substrate, which create a two dimensional
``egg-carton'' potential. The dynamics of the electrons in this
potential is described by a Hubbard model with long-range Coulomb
interactions. The state of the electrons in this system can be
probed via its conductance and noise properties. This allows the
identification of a metallic or insulating state. Numerical
estimates for the parameters appearing in the effective Hubbard
model are calculated using the proposed experimental system. These
calculations suggest that observations of quantum phase transition
phenomena of the electrons in the potential array are within
experimental reach.
\end{abstract}

\maketitle

Since the pioneering experiments of Greiner {\it et al.}, where
bosonic atoms were trapped in optical lattices to create a
Bose-Hubbard model \cite{greiner02}, quantum simulators have
attracted a large amount of attention. The interest is twofold.
The first is that such devices are hoped to offer an alternative
method for studying quantum many-body systems, which appear in
many branches of physics, yet in many cases no efficient and
reliable means are available to obtain quantitative information
about them. The second is from the perspective of quantum
computing, where the technology developed for such devices are
hoped to be useful for realizing a scalable quantum computer.
Indeed, many proposals for using optical lattices in quantum
information processing devices exist \cite{pachos03}.

In this paper, we propose another type of quantum simulator that
may be used for studying fermionic particles interacting via a
Hubbard-type interaction. The Fermi-Hubbard model, as opposed to
the Bose-Hubbard model, is particularly interesting due to its
significance in condensed matter theory, ranging from its original
inception in understanding metal-insulator transitions, to its
recent reincarnation in high-temperature superconductivity. In
particular, it is still controversial in whether the Hubbard model
supports a $d$-wave superconducting phase \cite{moriya03}. It is
also the basis for gate operations in electron spin-based quantum
computing \cite{Loss}. Recently there have been
many advances in the trapping of fermionic atoms in optical lattices \cite{kohl05}.
One of the features of the optical lattice approach is that
the effective form of the interactions is a
contact interaction that is extremely short ranged \cite{dalfovo99}. Since in a
real system displaying Hubbard dynamics the interactions are of a
long range form originating from the Coulomb interaction,
such effects may be difficult to be taken into
account in an optical lattice approach. One may speculate that to
simulate a Coulomb interaction, a real Coulomb interaction is
necessary in the system carrying out the simulation.

\begin{figure}
\label{setup}
\begin{center}
\hbox{\qquad\resizebox{6.0cm}{!}{\includegraphics{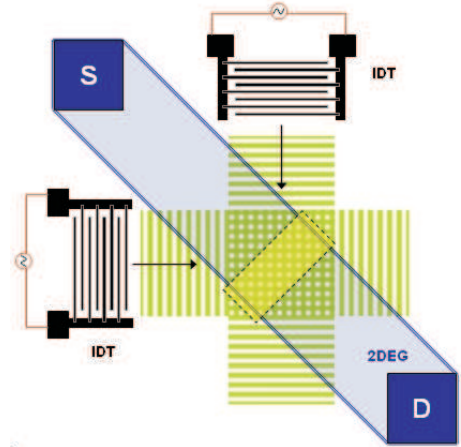}}}
\end{center}
\vskip -1cm \caption{ \label{fig:setup} (Color online) A sketch of
the proposed experimental structure. Two IDTs launch SAWs (green)
at right angles to each other forming a moving two-dimensional
``egg-carton'' potential. Light blue regions indicate the regions
where the 2DEG is present. Part of the 2DEG channel is shallow
etched (yellow), to partially deplete the electron density. Source (S)
and drain (D) ohmic contacts (dark blue) are formed at the ends of the
2DEG. }
\end{figure}
Our proposed experimental setup is shown in Fig. \ref{fig:setup}.
We start with a standard modulation-doped GaAs/AlGaAs
heterojunction, which forms a two dimensional electron gas (2DEG).
The 2DEG is formed into a channel as shown in the figure, with
source and drain ohmic contacts placed at the two ends. Between
the source and drain contacts, a region of low electron density is
formed via locally raising the conduction band, which may be done by 
a shallow etching procedure. The ohmic contacts thus probe the conductance properties
of this region. Outside of the 2DEG mesa,
interdigitated transducers (IDTs) are placed at right angles to
each other. Surface acoustic waves (SAWs) are launched by applying
a high frequency AC voltage to the IDTs, forming an interference
region at the center of the device. Due to the piezoelectric
property of GaAs, an electric potential is induced following
the shape of the SAW modulation. The net potential due to the 
SAWs in the central region is then equal to the sum of the
individual SAW potentials.  Assuming both IDTs produce identical SAWs, the total potential at a particular time is then
\begin{equation}
\label{eggcarton} V(x,y) = V_0 \left( \cos (k x) +
\cos (k y) \right),
\end{equation}
where $ V_0 $ is the amplitude generated by the
SAWs, $ k =2\pi f/v_{\mbox{\tiny SAW}} $ is
the wavenumber of the SAWs, with $ v_{\mbox{\tiny SAW}} $ being the SAW velocity ($ v_{\mbox{\tiny SAW}}
\approx 2860$ m/s in GaAs), and $ f $ the SAW frequency. This is an ``egg-carton'' shaped
potential that moves in the diagonal direction towards the drain electrode. We assume that the
approximate size of the interfering region that can be produced is
$50$ $\mu$m$\times 50$ $\mu$m. The depth of the potential can be
controlled by varying the amplitude $ V_0 $ of the
SAWs.

Such a moving potential may be used to transport electrons from the source to 
drain electrodes in a similar way to that observed in Ref. \cite{shilton96}. 
As the SAW moves, it
carries the trapped electrons with it in the local minima of the SAW (the acoustoelectrical
current).  If we assume that $ n $ electrons are trapped in each
minima, the acoustoelectrical current is then given by $ I = nef
$, where $ e $ is the electron charge. A crucial difference between the current setup and the
experiment of Ref. \cite{shilton96} is that the length of the
region between the source and drain electrodes was only of the
order of a single SAW wavelength in Ref. \cite{shilton96}.
Therefore, only one or two potential traps are realized in the
central region in their case. Assuming a SAW frequency of $f
\approx 20$ GHz in both directions (corresponding to a wavelength
of $\lambda = 0.14 $ $\mu\mbox{m}$), we have $\sim 10^5$ potential
minima in the central region. The effective lattice size will be
governed by the coherence length of the sample. Long
coherence lengths are now achievable due to advances in modern
fabrication techniques such that large sections of the lattice
will interact coherently. For example, mobilities $ \mu $ corresponding to
mean free paths of $ l_{\mbox{\tiny mfp}}= v_F \mu m^*/e \approx 11 $ $\mu\mbox{m}$ at T=0.1 K and
electron densities of $ n_d =  10^{10} $ $\mbox{cm}^{-2} $
were reported in Ref. \cite{umansky97}, where $ v_F $ is the Fermi velocity ($v_F \approx 2 \times 10^7 $ $\mbox{cm}/\mbox{s} $ in GaAs), 
and $ m^* $ is the electron effective mass ($ m^* = 0.067 m_e $ in GaAs). The properties of the
current flowing from the source to the drain are therefore
dependent on the collective effect of the lattice of electrons
formed by the SAWs.

To model the electrons in this interference region, let us start
with the general Hamiltonian
\begin{eqnarray}
\cal{H} & = & \sum_\sigma \int d^3 x \psi_\sigma^\dagger (\bm{x}) \Big[ - \frac{\hbar^2}{2m^*} \nabla^2 +  V(x,y) + V^z (z) \Big] \psi_\sigma (\bm{x}) \nonumber \\
& + & \frac{1}{2}  \sum_{\sigma \sigma'} \int d^3 x d^3 x'  \psi_{\sigma'}^\dagger (\bm{x}') \psi_\sigma^\dagger (\bm{x})
U_C (\bm{x},\bm{x}')  \psi_\sigma (\bm{x}) \psi_{\sigma'}
(\bm{x}'), \nonumber
\end{eqnarray}
where $V^z (z) $ is
a triangular potential well associated with the heterojunction,
and $ \psi_\sigma (\bm{x}) $ are the fermion field
operators for electrons of spin $ \sigma $. The Coulomb potential
$ U_C (\bm{x},\bm{x}') $ experienced by the electrons in the 2DEG
will be strongly screened due to the high density of electrons in
the $\delta$-doping layer located near the 2DEG. These electrons
are separated from the heterojunction by a distance $ d $, which may be
controlled by the thickness of the spacer layer. We therefore take
the form of the Coulomb potential to be
\begin{eqnarray}
\label{screenedcoulomb}
& & U_C (\bm{x},\bm{x}')  =  \frac{e^2}{4 \pi \epsilon}  \frac{1}{|\bm{x}-\bm{x}'|}  \\
& & \times \left[1- \frac{|\bm{x}-\bm{x}'|}{\sqrt{(x-x')^2 +
(y-y')^2+(z+z'+2d)^2}} \right], \nonumber
\end{eqnarray}
where $ \epsilon $ is the permittivity ($\epsilon \approx 13
\epsilon_0 $ in GaAs). Equation (\ref{screenedcoulomb}) was
derived assuming a metal plate in the $x$-$y$ plane a distance $ d
$ from the electrons.

The single particle Bloch wavefunction solutions of $ \cal{H} $ are simply Mathieu functions, 
from which we may construct a Wannier basis
\cite{wannier47,jaksch98}. There is a localized Wannier function $
w (\bm{x} - \bm{x}_j) $ for every local minima in the $x$-$y$
plane, labeled by the index $ j $. Making the transformation $
\psi_\sigma (\bm{x}) = \sum_j c_{j \sigma} w (\bm{x} - \bm{x}_j)
$, where $ c_{j \sigma} $ is a fermion annihilation operator
associated with the site $ j $, we obtain the Hamiltonian
\begin{eqnarray}
H & = & \sum_\sigma \sum_{ij} t_{ij} \left( c_{i \sigma}^\dagger
c_{j\sigma} + c_{j \sigma}^\dagger c_{i \sigma} \right)
+ \mu \sum_\sigma \sum_{i} n_{i \sigma} \nonumber \\
& & + \frac{1}{2} \sum_{\sigma \sigma'} \sum_{ijkl} {\cal
U}_{ijkl} c_{i \sigma'}^\dagger c_{j \sigma}^\dagger c_{k \sigma}
c_{l \sigma'} \label{latticehamiltonian}
\end{eqnarray}
where $ n_{i \sigma} = c_{i \sigma}^\dagger c_{i\sigma} $ and
\begin{eqnarray}
\label{tintegral}
t_{ij} & = & \int d^3 x w^* (\bm{x} - \bm{x}_i) \Big[ - \frac{\hbar^2}{2m^*} \nabla^2 +  V(x,y)  \Big] w (\bm{x} - \bm{x}_j)  \nonumber \\
\\
\label{uintegral} {\cal U}_{ijkl} & = &  \int d^3 x  d^3 x'
w^* (\bm{x}' - \bm{x}_i) w^* (\bm{x} - \bm{x}_j) U_C (\bm{x},\bm{x}') \nonumber \\ & &
\times w (\bm{x} - \bm{x}_k) w (\bm{x}' - \bm{x}_l)
.
\end{eqnarray}
If one chooses to discard all but the largest matrix elements in
Eqs.~(\ref{tintegral}) and (\ref{uintegral}), one obtains the
standard Hubbard model (in standard notation, $ t = -t_{ij} $, for
$ i,j $ nearest neighbors and $ U = {\cal U}_{iiii} $).
It is an important feature here that the full effect of
the Coulomb interaction is present in the Hamiltonian
(\ref{latticehamiltonian}), including long-range interactions and
correlated hopping terms. These are difficult to take into account
from a calculational point of view, but are naturally included in
the current setting.

\begin{figure}
\scalebox{0.65}{\includegraphics{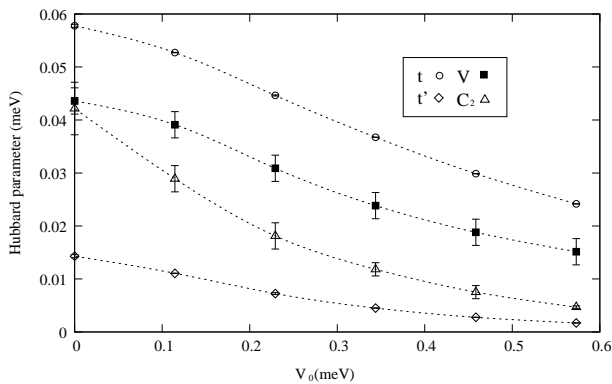}}
\caption{\label{fig:tVCplot}
 Various Hubbard parameters as a function of the SAW potential height.
$t$ is the nearest neighbor hopping, $t'$ is the next-nearest neighbor hopping,
$ V = {\cal U}_{ijji} $ is the nearest neighbor Coulomb integral, $ C_2 =  {\cal U}_{iijj} $ is a
pair correlated hopping term, with $i,j$ nearest neighbor sites (see inset of Fig. \ref{fig:uplot}).
The lattice sites involved in the various terms
are shown in the inset of Fig.~\ref{fig:uplot}. The chosen
parameters are a SAW frequency of $ f = 20$ GHz and a screening
distance of $ d = 10$ nm. Effective
mass and permittivity values were taken to be those in GaAs. Error
bars are due to the Monte Carlo integration procedure used. Lines
are to guide the eye. }
\end{figure}

\begin{figure}
\scalebox{0.65}{\includegraphics{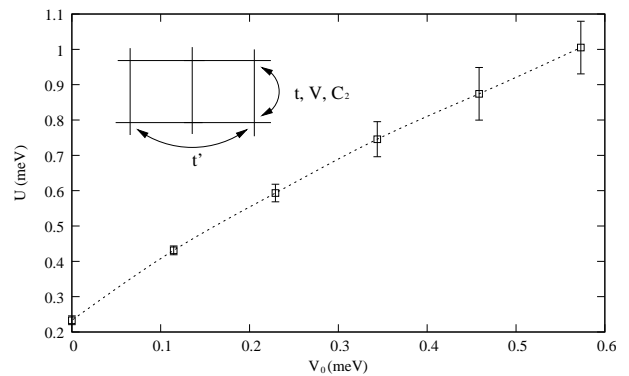}}
\caption{\label{fig:uplot} The on-site Coulomb integral as a
function of the SAW potential height. The inset shows the
connecting points for the various Hubbard parameters calculated in
Fig.~\ref{fig:tVCplot}. Lines are to guide the eye. }
\end{figure}

Let us evaluate the integrals Eqs.~(\ref{tintegral}) and
(\ref{uintegral}) for the proposed experimental parameters. The
results are shown in Figs.~\ref{fig:tVCplot} and \ref{fig:uplot}.
First, we find that the only hopping directions that are allowed
are those along the lattice axis directions, i.e. all diagonal
hoppings are zero. Therefore, the next-nearest neighbor hopping
term $ t' $ is a two lattice site hopping term as shown in the
inset of Fig.~\ref{fig:uplot}.  This is simply a result of the
Wannier basis that was chosen to write the Hamiltonian
Eq.~(\ref{latticehamiltonian}) and is due to the symmetry of the
square lattice that is currently being used.  All hopping terms
decay with a roughly exponential dependence on the SAW potential
height $ V_0 $.  The on-site Coulomb term $U$ increases with $V_0$
as expected due to the greater localization of the Wannier
functions for larger potentials. The nearest-neighbor Coulomb term
$V$ (i.e. the coefficient of the term $ n_{i \sigma} n_{j \sigma'}
$ with $i,j $ nearest neighbors) approaches the asymptotic value
of $ V(V_0 \rightarrow \infty) = \frac{e^2}{4 \pi \epsilon
\lambda}(1-1/\sqrt{1+4d^2/\lambda^2}) \approx$ 8 $\mu$eV,
consistent with the numerical calculation. We also calculate one
of the correlated hopping terms denoted as $C_{2}$ in Fig.~2 which
is the coefficient of the term $ c_{j \uparrow}^\dagger c_{j
\downarrow}^\dagger c_{i \uparrow} c_{i \downarrow} $ where $i,j$
are nearest neighbor sites. Due to the small amplitudes many of
the correlated hopping terms are numerically difficult to
calculate. This correlated hopping term appears to have the
largest amplitude of these.

\begin{figure}
\scalebox{0.8}{\includegraphics{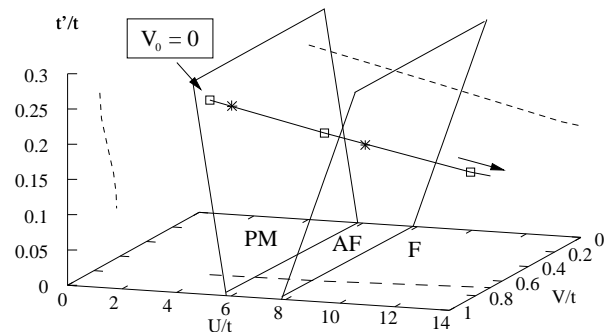}}
\caption{\label{fig:trajectory}  The trajectory in the space
$(U/t,V/t,t'/t)$ as the SAW potential is increased. The phase
boundary between the paramagnetic metal (PM) and antiferromagnetic
(AF) regions, as well as the boundary between the
antiferromagnetic and ferromagnetic (F) regions (both insulating)
as obtained from mean field theory at filling $n=1.3$ is also
shown. The stars indicate the points where the parameters pass
through the boundaries. Dashed lines show the projection of the
trajectory onto the planes spanned by the $(U/t,V/t,t'/t)$ axes.
 }
\end{figure}

One expects that due to the Hubbard nature of the electron
interactions in Eq.~(\ref{latticehamiltonian}) that as the SAW
potential is raised, at some point a metal-Mott insulator
transition will be reached  \cite{eisensteincomment}. Figure \ref{fig:trajectory} shows the
trajectory of the Hubbard parameters in the space $(U/t,V/t,t'/t)$
as the SAW potential is increased. On this trajectory we have
superimposed a magnetic phase transition boundary in this
parameter space as obtained from mean field theory, following
Refs. \cite{lin87,chattopadhyay97}. We use the Hamiltonian
Eq.~(\ref{latticehamiltonian}), with all parameters other than $U,
V, t'$ and $ t $ set to zero. Self-consistent solutions are found
for paramagnetic, antiferromagnetic, ferromagnetic, and charge
density wave states, and the phase diagram is mapped by taking the
lowest energy solution. We see that the system crosses over from a
paramagnetic metal phase to an antiferromagnetic phase,
then to a ferromagnetic phase as the SAW potential is
increased. We find that the effect of the two-site hopping $t'$ is
qualitatively similar to the more conventional diagonal hopping
term, with good agreement between our results and those obtained
by a diagonal coupling given in Ref. \cite{lin87}. Although it is
unclear whether the paramagnetic to antiferromagnetic transition
as calculated in mean field theory corresponds precisely to the Mott
transition point, our results suggest that our parameter
values are in the correct range to observe quantum phase
transition phenomena.

\begin{figure}
\label{bandstructure}
\begin{center}
\hbox{\qquad\resizebox{6.5cm}{!}{\includegraphics{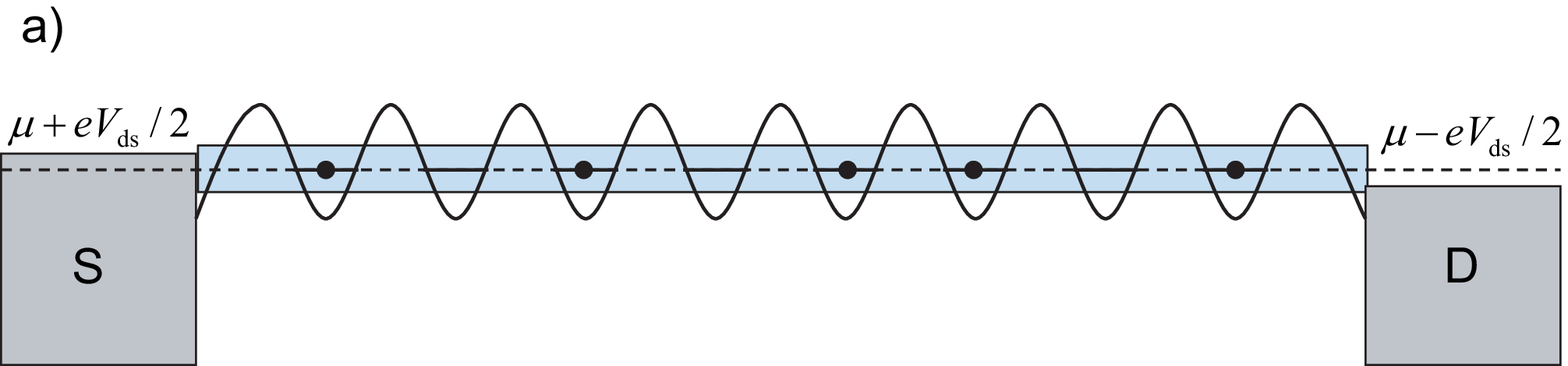}}}
\end{center}
\begin{center}
\hbox{\qquad\resizebox{6.5cm}{!}{\includegraphics{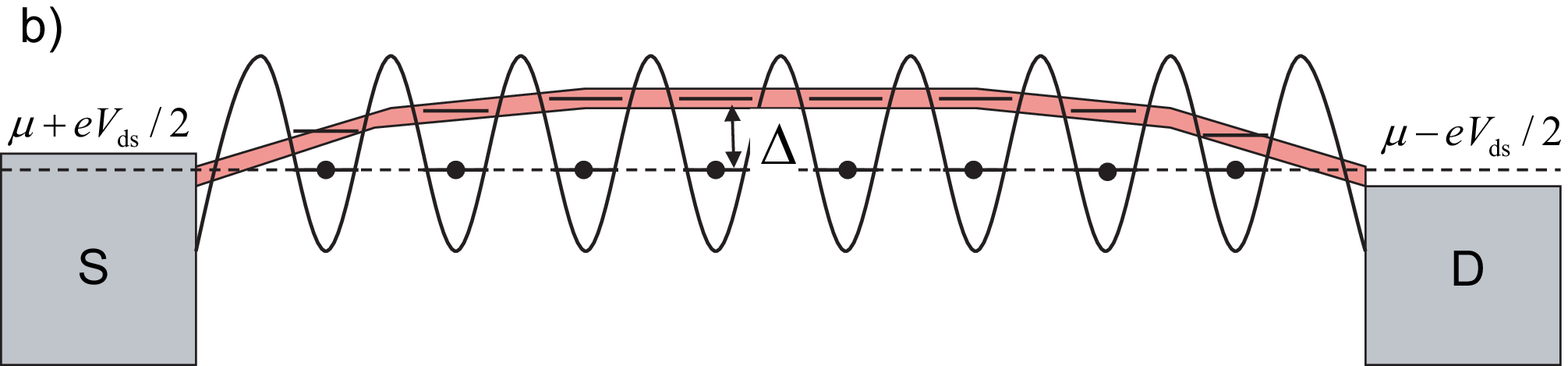}}}
\end{center}
 \caption{ \label{fig:band} (Color online) Band
structure (cut along the diagonal in Fig.~\ref{fig:setup} of the
Hubbard simulator coupled to source (S) and drain (D) reservoirs.
The metallic phase is depicted in a). The energy to add or remove
electrons via the source and drain electrodes is gapless which
means that the Fermi level $\mu$ (dashed line)  of the reservoirs
lies within the conduction band (blue).
%of the 2D lattice region that evolves out of the single
%well quasi bound states.
The Mott-insulator phase is shown in b) for filling $ n = 1 $.
Changing the number of electrons in the lattice region is
associated with a charge gap $\Delta$ which suppresses
dc-transport if $k_{B}T,eV_{\rm ds}<\Delta$. This is indicated by
a lifted conduction band (red) in the bulk of the lattice.}
\end{figure}

This transition will be reflected in a modified band structure in
the central region of the setup shown in Fig.~\ref{fig:setup}.  In
Fig.~\ref{fig:band} we show the band structures due to two SAW
intensities corresponding to the metallic and Mott insulating
phases. First consider the weak SAW modulation case
(Fig.~\ref{fig:band}a). The periodic potential
Eq.~(\ref{eggcarton}) produces a conduction band as shown in the
figure. Due to the weak modulation, the band structure corresponds
to a metallic state, i.e. there is no gap between the ground
states of successive electron numbers. The band structure due to
the strong SAW modulation is shown in Fig.~\ref{fig:band}b. Ground
states with successive electron numbers ${N}=\sum_{i}\langle {\hat
n}_{i}\rangle$ are separated from each other by a charge gap $
\Delta $. Now consider applying a small source-drain voltage
$V_{\rm ds} $. In general this will add a dc-current $I_{\rm dc}$ component
to the acoustoelectrical current. 
This dc-current probes the excitations or the phase of
the Hubbard model. Considering the Mott insulator regime
(Fig.~\ref{fig:band}b), when a small bias voltage $V_{\rm ds}$
such that $eV_{\rm ds}, k_{B}T<\Delta$ is applied, the dc-current
is blocked as a consequence of strong correlations. On the other
hand, in the metallic phase no gap opens and Ohm's law should hold
$I_{\rm dc}\propto V_{\rm ds}$ (Fig.~\ref{fig:band}a). The phase
transition is reached by tuning the 2DEG electron density $n=N/A$
where $A$ is the area of the SAW lattice. The density can be
controlled by a backgate voltage and/or the SAW potential strength
$V_{0}$.

The absence or existence of fluctuations in the occupation numbers
${\hat n}_{i}$ can be probed by low-frequency shot noise. The shot
noise due to $I_{\rm dc}$ is supressed in a macroscopic conductor
\cite{Liu}. However, in an insulating phase the fluctuations in the SAW driven acoustoelectrical
current
should still be observed due to the quantized bunches
of electrons per SAW-potential minima. 
Assuming zero, one, or two electrons per site, 
the shot noise is proportional to
$f\langle(\delta {\hat n}_{i})^{2}\rangle$ with $\langle(\delta
{\hat n}_{i})^{2}\rangle=\langle{\hat n}_{i}\rangle(1-\langle{\hat
n}_{i}\rangle)+2\langle{\hat n}_{i\uparrow}{\hat
n}_{i\downarrow}\rangle$ where ${\hat n}_{i}={\hat
n}_{i\uparrow}+{\hat n}_{i\downarrow}$ \cite{thnoise,expnoise}. 
The complete absence of doubly occupied and empty sites is only predicted at half-filling
in the limit of infinite $ U $ \cite{mancini00}. 
Therefore, at half-filling and deep into the Mott insulator regime the insulating phase
is characterized by the disappearance of shot noise. Off half-filling, 
the electron number on a given site is free to fluctuate and we expect a finite shot noise. 
 Noise measurements in combination
with conduction measurements therefore can clarify the role of fluctuations in the occupation
numbers in the insulating regime. 

In order to observe a quantum phase transition, we require
$ k_B T $ to be smaller than the typical energy scale of the
Hamiltonian, which is set by the hopping integral $ t $
\cite{sachdev99}. We expect that a phase transition will occur for
relatively small values of the SAW potential, where $ t \approx
0.05 $ meV. This corresponds to temperatures less than $ T \approx
0.6$ K, which is an experimentally accessible temperature.
Another consideration is due to the fact that we are using
a moving potential which remains in the central region of
Fig.~\ref{fig:setup} for a time of approximately $L/v_{\mbox{\tiny
SAW}} $, where $ L $ is the length of the central region.
This time corresponds to approximately
15 ns for $L\sim50$ $\mu$m, which must be larger than the
tunneling time for the electrons in the lattice. As the tunneling
time is of the order $ \hbar/t \sim 10$ ps, there is time for
many tunneling events and therefore equilibration should not be
problematic in this case.

A simple variation on the current proposal is to independently vary the SAW potential
heights in the $x$ and $y$ directions,  which changes the hopping amplitudes
$ t_{ij}$ in these directions. By making the SAW potential much larger in one direction, 
we may smoothly evolve the system 
from a 2D lattice to a 1D chain. Another possible variation 
is to add a random
potential $R(x,y,t)$ to the potential Eq.~(\ref{eggcarton}). This is
easily implemented by applying a random amplitude modulation voltage
to the IDTs. The
effect of such a term is to randomly distribute both the chemical
potential $ \mu $ and hoppings $ t_{ij}$ from site to site in the
Hamiltonian Eq.~(\ref{latticehamiltonian}). This is a
Hubbard-Anderson model and has been considered in studies of
interacting disordered systems \cite{belitz94}. The random
potential can be characterized by two parameters, the standard
deviation of the amplitude $ \sigma_R $ which controls the degree
of disorder, and the correlation $ \int R(x+\Delta x, y+ \Delta
y,t) R(x,y,t) dx dy $ which controls the density of the
``impurities''. By simultaneously exciting higher harmonics of the
IDT, and length modulating the amplitudes of these harmonics
randomly, the correlations of $R(x,y,t)$ may be varied. In this
way we may study metal-insulator transitions brought on by the
effects of disorder, as well as correlations.

This work is supported by JST/SORST, NTT, and the University of
Tokyo. T. B. is supported by a JSPS fellowship. We would like
to thank Y. Hirayama, T. Fujisawa, and S. Sasaki for helpful discussions.

%%%%%%%%%%%%%%%%%%%%%%%%%%%%%%%%


\begin{references}
\bibitem{eisensteincomment}
We note that a metal-insulator transition in a 2DEG as a function of the electron density
was observed in Ref. \cite{tracy05}. The type of transition discussed here differs to that of 
Ref. \cite{tracy05} in that we assume a higher electron density ($n_d \gtrsim 10^{10}$ $\mbox{cm}^{-2} $) to screen 
out effects due to an inhomogeneous potential in the 2DEG plane, which localizes electrons in Ref. \cite{tracy05}.
\end{references}
\end{document}